\documentstyle{article}
\begin{document}

\title{Big Bang Nucleosynthesis and the observed abundances of  light elements}

\author{Craig J. Hogan\\
University of Washington}

\maketitle

\section{Introduction}

The standard model of the early universe is
extraordinarily simple: it just assumes a globally
isotropic and  uniform universe.
In the simplest version there is no structure
of any kind on scales larger than individual elementary particles;
indeed the  contents are determined  by
``just physics''--- global expansion governed by
relativity, particle interactions governed by the Standard
Model, and  distributions governed by statistical
mechanics. Since the gravity is itself dominated by the
relavistic plasma early on, parameters that become prominent
at late times (global curvature, dark matter density) make no difference,
so the model has  only one  parameter: the ratio  $\eta$
of the number of baryons to the number of photons,
often expressed in units of $\eta_{10}\equiv \eta/10^{-10}$.
Even this parameter   has almost no effect on the
early evolution of the universe,  but it does affect the
principal observational relic of the early expansion, the
  relative abundances
of light nuclei.
Since the number of photons
in the universe is known today from the microwave background temperature
(it is $411\pm 2 cm^{-3}$), $\eta$ specifies the present day baryon density:
$\Omega_bh^2=3.65\times 10^{-3}\eta_{10}$.

The model would be beautiful even if simplicity were
its only virtue, but remarkably,  as observations improve they verify
it   with increasing precision as a good description of the real world.
The most precise verification of the basic framework
 now comes from observations of the
   isotropy  and spectrum of the microwave background, which verify the precise
uniformity of the universe on large scales the primordial origin of the
 radiation.  Here I discuss the other principal observational
relic of the early universe, the cosmic abundances
of light nuclei, which  probe   homogeneity over a much wider
range of scales than the microwave background, and which record
processes starting with  redshifts of the order of $10^{10}$.
In particular, I discuss 
 observations of   abundances as a precise test of the model and
a precise test of the parameter $\eta$.

Unfortunately the later evolution of the universe is anything but simple.
The uniform gas of the Big Bang long ago converted into
a complex universe of stars and gas, which has undergone considerable
nuclear evolution. The main problem is
to connect what we can actually observe today with the very clean predictions
of the simple Big Bang
model, measuring abundances in places where
we can uncover fairly well-preserved relics of the initial abundances
or deduce constraints on them.

\section{The Model Predictions}

The nuclear and statistical physics of Big Bang nucleosynthesis are
thoroughly studied and its
predictions and  errors are well characterised using
  numerical  integration of reaction networks to predict the
abundances for the light elements ( \cite{pje66},\cite{wfh67}, \cite{wag73}).
The predictions can be shown graphically in the traditional
plot as functions of $\eta$ (e.g.\cite{wal91},\cite{smi93},\cite{cop95a},
 \cite{steigman96},
).
Also useful are  the following
fitting formulas
(adapted from from \cite{sar96}),
including $1\sigma$ theoretical errors estimated by Monte Carlo techniques.
The predicted fraction of total baryon mass in helium-4 is given
 by
$$
Y_P=0.235+0.012\ln\left({\eta_{10}\over 2 }\right)
\left({\eta_{10}\over 2}\right)^{-0.2}
+0.011\left[1-\left({\eta_{10}\over 2}\right)^{-0.2}\right]
\pm 0.0006
$$
with theoretical error
 due in about equal parts to uncertainty in the neutron lifetime
and in nuclear reaction rates.
The abundance by number of  deuterium is
$$
\left({D\over H}\right)_P
=15.6\times 10^{-5\pm 0.03}\left({\eta_{10}\over 2}\right)^{-1.6},
$$
or
$$
\log \left({D\over H}\right)_P =-3.81-1.6\log \left({\eta_{10}\over 2}\right)
\pm 0.03
$$
with errors dominated by nuclear rates.
 The abundance of
 lithium-7 is
$$
\left({^7Li\over H}\right)_P
=1.06\times 10^{-10\pm 0.1}\left[
\left({\eta_{10}\over 2}\right)^{-2.38}
+0.28\left({\eta_{10}\over 2}\right)^{+2.38}\right],
$$
with   theoretical error again due to reaction rate
uncertainties.
The Big Bang produces  negligible amounts of anything heavier,
and we omit helium-3 from this discussion
 because its primordial abundance is not
measurable.

  It is a considerable   challenge 
to measure the actual primordial abundances to a comparable level of precision,
both because of uncertainties in measuring present-day abundances and
because of uncertainties in modeling the nuclear evolution since
the Big Bang. For each of these nuclei however there is a favorite 
place to look.

\section{Helium in extragalactic HII regions}

Nuclear evolution after the Big Bang always increases
the abundance of helium, usually but not always accompanied
by production of heavy elements. At present
the primordial abundance of helium-4 is best estimated in hot, low-metal
HII regions in low redshift galaxies. The plasma is hot and
largely photoionized by hard radiation from young stars.
Abundances are estimated from the strengths of the emission
lines of  hydrogen and helium. Since both are recombining
from the same electron distribution, this gives a fairly direct
abundance estimate from the relative recombination
and transition rates, with only small corrections due to small
differential variations with temperature, line
radiation transfer, and collisional excitation.
In the best studied cases $Y$ is measured with an accuracy
of a few percent; for example, in IZw18, Skillman and Kennicutt\cite{sk93}
find $Y=0.231\pm 0.006$. Several dozen such regions have been
measured with useful accuracy. Broadly speaking, they reveal
the unmistakable signature of the Big Bang: a universal minimum
 abundance of helium,
with $Y_P\approx 0.23$.

The sample is large enough to
attempt a more precise estimate of
$Y_P$  from the set of estimated $Y$'s in various
nebulae.
A widely adopted technique   is to measure the abundance
of another element-- such as O or N-- in several regions,
then extrapolate to zero metallicity to estimate the primordial value.
This also gives some information about the enrichment history
of the gas. A linear relation is    not however necessarily expected
(especially in such small regions), so one should entertain
the possibility that the enrichment is stochastic in character.
In this case one should
simply take the lowest values, or some set of lowest, best
measured values, taking care to avoid a statistical bias
in the estimate of $Y_P$. The final estimate of $Y_P$ turns
out to be remarkably insensitive to which assumptions are made
or which subsamples are used.
Recent independent analyses yield:
$Y_p=0.228\pm 0.005\pm 0.005$ (\cite{pag92}),
$Y_p=0.231\pm 0.006$ (\cite{sk93},\cite{ski93}),
 $Y_p=0.232\pm 0.003\pm 0.005$ (\cite{oli95}),
$Y_P=0.230\pm0.006\pm 0.004$ (\cite{skill96}), and
$Y_p=0.234\pm 0.005$ (\cite{pei96}). The fits are dominated
by a small number of galaxies  (less than 10) which are well
measured and show low values of $Y$.
A significantly higher value of  $Y_p=0.243\pm 0.003$
is derived by Izotov et al (\cite{iz94}, \cite{izo96}), largely because of a different
sample which excludes many of these galaxies.

The  systematic error of the order of 0.005 in all of these
is mainly due to uncertainties in modeling the HII regions, for example,
the amount of neutral helium and the  collisional excitation of $HI$,
which tend to cause underestimates of $Y$,
and temperature fluctuations, which tend to cause overestimates.
The range of estimates reflects these uncertainties.
It will be hard to reduce these errors significantly,
but they are well controlled at the 0.005 level.  (In particular the
uncertainty
due to calculated emissivities of helium lines
 have largely disappeared.)

From the formulas above, an error in $Y_P$ propagates to an
error in $\eta$ via
$$
 (\delta\eta/\eta)\approx 83 \delta Y_P    
$$
so $\eta$ is poorly constrained by $Y_P$, to no better than
a factor of two. On the other hand,
the helium observations are a powerful confirmation of
the Big Bang picture since the prediction is so insensitive
to the model parameter. The helium observations are for example
clearly consistent with the number of baryons actually observed,
which correspond to about $\eta_{10}\approx 2$ (see below).

\section{Deuterium in Quasar Absorbers}

The nuclear evolution of deuterium is the opposite of
helium. It is only made in the Big Bang, and stellar
processing always decreases its mean abundance\cite{ep73}.
Except for local enhancements in  molecular
forms  (due to fractionation), the highest
reliable abundance gives a lower limit on the primordial
value and an upper limit on $\eta$. It is quite sensitive
to $\eta$ which makes it the best tool for a precise measurement.

The most promising technique is to estimate the
deuterium  abundance    from
quasar absorption lines.
They provide a census of material in a wide range of
environments over a huge volume of space, in particular including
high redshift material which is relatively unprocessed.
They also provide in principle a precise estimate of abundance
free of many complex astrophysical effects, since the column densities
of two species ($DI$ and $HI$) are related in a simple way to the
absolute abundance in the  gas (charge exchange reactions for example
guarantee that the ionization is nearly identical for the two species),
and both column densities in some situations can be accurately estimated
from optically thin or damped absorption in Lyman series lines.

In practice the situation is not yet quite so clean.
The holy grail in the field is an absorption system with very simple
velocity structure (a single isolated cold cloud is best,
with a temperature low enough to make the $ D$ identification
secure),
with a flow field and temperature highly
constrained by metal lines (but still with low metal abundance,
so that $D$ is not destroyed
significantly), with high S/N, high resolution data
including well resolved,
optically thin Lyman series lines and extending
well beyond the Lyman limit. No system yet combines all of these
attributes, although some come close.

 Current results are listed in the accompanying table. Note that the errors
quoted are not total errors, but are just the errors from column
density parameters in a model fit. They do indicate the
precision the technique is realistically
 capable of, and the total error could be as small
as this in a favorable situation.

The best absorber yet found for this purpose is the $z=3.32$
 system in Q0014,
where accurate columns can be measured for both the deuterium
(from Lyman $\alpha$) and the hydrogen (from high order Lyman
series lines). The formal error is only about 25\% in this measurement.
There is a   nonnegligible probability that the D$\alpha$ line is
not a deuterium feature at all but a hydrogen line
at just the velocity where it masquerades as deuterium;
  such contamination is rare enough
that it  does not significantly affect
either the value or the error, although it introduces a small
upward bias and changes the confidence intervals considerably.

In this and another absorber in the same quasar,
 we have found evidence (\cite{rug96a},\cite{rugers96})that the
deuterium lines are narrow, as expected for a deuterium feature,
but unusually narrow for hydrogen.  This supports the interpretation
of the feature as deuterium since it reduces the probability
of an incorrect identification
 to less than $10^{-3}$.  There is still the possibility that
$HI$ systems of high column tend to be associated with
cold, low column $HI$ companion clouds which mimic deuterium and
increase the probability of a spurious identification; this possibility
we are checking with hydrodynamical cosmological simulations
and control samples.

\bigskip

\centerline{Published and Reported $D/H$ from Quasar Lyman Series
Absorption}
\bigskip\bigskip
\def\tablerule{\noalign{\hrule}}
\halign to \hsize{#\hfil\tabskip=1em plus
.5em&\hfil#&#\hfil&#\hfil&#\hfil\tabskip=0em\cr
log D/H &Quasar &z &Comments &Reference\cr
\noalign{\vskip 4pt}
\tablerule
\noalign{\vskip 6pt}
$-3.66 \pm 0.06$ &0014+813 &3.32 &Ly$\alpha$-Ly 17 &\cite{schr}\cr
$-3.6 \pm 0.3$ &&&&\cite{car94}\cr
$-3.72 \pm 0.1$ &&&narrow D$\alpha$, Ly$\alpha$-Ly$\mu$ &\cite{rug96a}\cr
\noalign{\vskip 12pt}
$-3.73 \pm 0.28$ &0014+813 &2.8 &narrow D$\alpha$, Ly$\alpha$, metals
&\cite{rugers96}\cr
\noalign{\vskip 12pt}
$-3.9 \pm 0.4$ &0420-388 &3.086 &D$\alpha$-D$\gamma$, metals &\cite{car96}\cr
$-3.7 \pm 0.1$ &&&OI/DI = const. assumed &\cr
$>-4.7$ &&&conservative &\cr
\noalign{\vskip 12pt}
$\leq -3.82$ &1202-0725 &4.672 &D$\alpha$, metals, high O/H &\cite{wam96}\cr
\noalign{\vskip 12pt}
$-$4.64 $\pm$ 0.06 &1937-1009 &3.572 &D$\alpha$, Ly$\alpha$-Ly 17, metals
&\cite{tyt94}\cr
\noalign{\vskip 12pt}
$-$4.2 to $-$4.0 &1937-1009 &3.572 &fit to Tytler et al's model
&\cite{wampler96}\cr
\noalign{\vskip 12pt}
$-3.95 \pm 0.54$ &0636+680 &2.89 &D$\alpha$, Ly$\alpha$ only &\cite{rug96b}\cr
\noalign{\vskip 12pt}
%$-$3.7 &0956+122 &-- &D$\alpha$-- D$\gamma$ &Songaila \& Cowie\cr
%\noalign{\vskip 12pt}
$-$4.60 $\pm$ 0.08 &1009+2956 &2.504 &D$\alpha$, Ly$\alpha\beta\gamma$, metals
&\cite{bur96}\cr
\phantom{$-$4.60 }$\pm$ 0.06 &&&&\cr
\noalign{\vskip 6pt}
\tablerule}
\bigskip
\bigskip

The pattern emerging from most of these data is that of a   ceiling
at around $\log(D/H)\approx -3.7$ or $(D/H)\approx 2\times 10^{-4}$; no
measurements are found
significantly above
this value, while several accurate estimates  lie close to it. This
is of course the pattern expected in the model where this is about
the primordial value, which is subsequently reduced in patches
by stellar processing.

Two of the measurements lie well below the others, leading to a
dichotomous situation reminiscent of the former situation
with the Hubble constant. This dichotomy can however probably be resolved
soon, as either genuine or due to observational artifacts.

If the differences turn out to be genuine, the most likely
explanation will be that the
  low values
are caused by patchy D destruction. On the scale of
the absorption clouds D destruction  does not necessarily correlate
in microscopic
detail with local metal abundances, since the two effects  are dominated
by stars of different masses, and the clouds may be small enough
(especially for low ionization parameter)
not to represent a fair sample of a stellar IMF.
For example, in a star formation region of less than about a thousand
solar masses, the expected number of massive stars is
small enough  even with a standard IMF to occasionally have
no supernovae at all, so the ejecta are both metal poor
and deuterium poor.
 A few rare
low  values of $D/H$  may therefore
be consistent with a high $(D/H)_P$ and low metals.

For example, in the 1937-1009 absorber fit by Tytler
et al \cite{tyt94}, there are two components with very different metallicities.
If one were to add a third component of dense
gas with low ionization parameter and high $HI$ column,
 with no detected metals
(hiding them in low ionization states), in which
the deuterium had been destroyed, the $D$ abundance of the other
two fitted components could be quite high. If the ionization parameter
is low enough, the neutral fraction of the extra component could be
high, allowing the extra cloud to be physically small so that
the D destruction is plausible, due say to a confluence of stellar winds
or a thermally unstable wind shock.

With small numbers of systems
we must also admit the possibility of errors interpreting
the observations, such as the possibility
 that the high $D/H$ values are erroneous due
to $H$ contamination. A high S/N
UV spectrum of the Lyman series of the 0014 absorbers could
disprove this possibility, since the
narrow  $D$  feature (with little turbulent
broadening)  predicts in detail the shape of
the high order $H$ lines.
On the other hand , the low $D/H$ values so far are
 both found in
systems where the $HI$ column is difficult to measure accurately because of
its high column density and highly saturated lines.
 Wampler has  argued recently  that the line spectra of $D$, $H$ and
several ion species  described by Tytler et al's
model in Q1937-1009 can be fit with three velocity components,
yielding an excellent fit with
 an $H$ column three times lower and $D/H$ three times higher than
the two component model. He conjectures that if the data were fit
instead of the model, an additional factor of two would be allowed
(depending primarily of the true level of flux allowed in the
troughs of the high order Lyman series lines),
bringing it up to the level of the high estimates, $D/H \approx 10^{-4}$.
This can only be true however if Tytler et al have significantly
underestimated the flux in the high order lines and the Lyman
limit, a possibility that will be tested soon with new
low resolution spectrophotometric data.

It is not difficult at present to reconcile the
  Q1009+2956 data \cite{bur96} with a high abundance, since the Keck data
in that system does
not yet extend to the blue far enough to separate
unsaturated components at high resolution;
the only good measure of total $HI$ column is from the Lyman limit, but this
does not provide a precise fix on where the $HI$ is in redshift (ie,
how much of it belongs to the deuterium). Here again, the issue
can be resolved with new data--- a high resolution UV spectrum.

There is a methodological question of how reliable the
whole technique
of Voigt profile fitting is for measuring abundances in QSO absorbers.
The feeling has been that it probably works well if there are enough
independent constraints, such as metal lines and multiple Lyman
series lines, and if a good fit is obtained with a fairly simple model.
We now have the opportunity to test this leap of faith quantitatively
since we have realistic hydrodynamical simulations of the clouds where
the ``ground truth'' is known;
these will
let us calibrate the accuracy of the profile fitting procedure
in naturally occurring clouds, and perhaps to identify those
situations where measurements are most reliable, including the
use of metal lines to guide the modeling of flows. Certainly
with a large enough sample of clouds to deal with   interlopers, this
is not an insuperable problem,  as long as the
gas is hot enough that the thermal part of the line
broadening is not much less than the spectral resolution
(which is true for photoionized $D$ and $H$ under
protogalactic conditions seen in HIRES spectra):
in this situation the atomic
 column of optically thin absorption
is measured directly and reliably from the optical depth.

It is important to enlarge the sample, both by enlarging the
search for a ``holy grail'' system to other sightlines, and by
lowering our standards and extracting information from
systems which are harder to interpret. For example,
not listed in the table are the recent estimates by Martin Rugers
and myself of $D/H$ in seven other systems of the Q0014 and Q0636 lines of
sight.
They all have large errors (most of them are based on
fitting of single line profiles)  and each individually
carries little weight;
nevertheless it is significant that no absorber was found
which required a  low abundance,    the distribution gives a formal
estimate of the mean consistent with the best measured systems,
and the distribution of Doppler parameters is consistent
with the deuterium identification.

\section{Deuterium and Helium-3 in the Galaxy}
In the Galaxy today, the interstellar $D/H= 1.5\pm 0.2\times 10^{-5}$
(\cite{McCullough}, \cite{lin95}, \cite{Linsky-etal}) or $1.3 \pm 0.4 \times
10^{-5}$\cite{fer96}. Although this is a lower limit to the primordial
abundance,
 we do not know the history
of the Galaxy well enough to reconstruct the primordial
abundance from Galactic observations.
If the primordial value were indeed ten times bigger,
 90\% of the deuterium
must have been destroyed   by now, or equivalently, only
10\% of the gas can be unprocessed.
  This happens in some chemical evolution models
which agree with the other constraints\cite{scully-etal96}.
More than 90\% of the disc mass is now
locked up in stars and remnants.
In successful models, massive, metal-producing stars
 and supernovae power early winds which eject much of the enriched gas;
 the
ISM  is then replenished by gas quiescently  ejected from
the envelopes of low mass stars, which has been depleted in deuterium.

Some estimates of $(D/H)_P$ are based on measurements in the solar
system--- not of D, but of $^3He$, which is what D is
turned into.  This is combined with stellar and Galactic
evolution models to derive a limit on $(D/H)_P$, based on the idea
that most stars do not destroy $^3He$,
and that therefore $(D/H)_P$ cannot be too large
without producing too much $(D+^3He)/H$ (\cite{ST92},
\cite{Dearborn-Steigman-Tosi}, \cite{Palla-etal}, \cite{Tosi}). However, we
reject
this argument as not reliable for at least two reasons:
(1) Observations of carbon, nitrogen and oxygen
 isotope anomalies in giant branch stars, and models
of giant branch mixing (``cool bottom processing'') inspired by them,
have cast doubt on the notion that low mass stars cannot destroy $^3He$;
processes that can change the isotopes also destroy $^3He$ by
large factors (\cite{hog95}, \cite{cha95}, \cite{was95}, \cite{weiss96},
\cite{booth}, \cite{booth96}). (2) Observations of interstellar $^3He $ in
hyperfine-structure emission   seem to show that stars can either make or
destroy $^3He$, and indeed the radial trend  in the Galaxy resembles
net destruction more than net production\cite{wil94}.
A handful of planetary nebulae\cite{rood92} show very high $^3He/H$
(much too high to be typical  \cite{gal96}), while   HII regions show
a large spread of values, including some very low ones which
are evidence of destruction\cite{wil94}.

 Both of these developments
argue for caution in using $^3He$ for anything other than a
probe of Galactic chemical evolution.
A growing consensus of opinion favors abandoning helium-3
as a cosmological probe altogether until its Galactic chemical evolution
is better understood. Not only is the  present abundance
difficult to measure outside the solar system,
 the primordial $D+^3He$ value is impossible to
deduce since the abundance has been altered significantly by
an unknown amount and unknown sign.

Recently, solar
system measurement of local interstellar ``pick-up ions''
(by the Ulysses spacecraft  \cite{glo96}) indicate that
the abundance today is $^3He/H= 2.1^{+0.9}_{-0.8}\times 10^{-5}$. The sum
of this and interstellar deuterium ($D/H=1.5\pm 0.2\times 10^{-5}$) is
$3.7\pm 0.9\times 10^{-5}$. For the presolar nebula, the guesses
from meteorite data are  $1.5\pm 0.2\pm 0.3 \times 10^{-5}$,
$2.7\pm 0.5\pm 1 \times 10^{-5}$ and $4.2\pm 0.7\pm 1\times 10^{-5}$,
respectively. These numbers are consistent with a steady conversion of
D into $^3He $, with little other $^3He $ production; they are also
consistent with a destruction of $D+^3He$ by a factor of two over the
last 4.5 Gy. These are interesting and important constraints
on Galactic chemical evolution, but do not contain primordial information.

\section{Lithium in Old Stars}

The primordial lithium abundance  estimated from samples
of metal-poor halo stars\cite{Spite-Spite}
is according to several estimates
$\log(Li/H)=-9.80\pm 0.16$ (\cite{wal91}, \cite{smi93}),
$\log(Li/H)=-9.78\pm 0.20$ (\cite{tho94}), log (li/H) = $-9.79$ to
$-9.76$\cite{Molaro-Primas-Bonifacio}.  Substantial uncertainties enter which
make lithium a less precise  cosmological tool than
either deuterium or helium.
 The SBBN predictions themselves are uncertain
because of uncertain reaction rates. The absolute abundance is uncertain
due to modeling of stellar atmospheres (which are more
reliable for differential measurements); and extracting primordial
abundances from the stellar abundance is perilous since the amount of
depletion is still controversial.
Abundances of  lithium in stellar atmospheres are influenced
significantly by settling and mixing, and by steady mass loss
and winds. Indeed all three of these processes are thought to
be significant in order to get even approximately a plateau independent
of temperature as observed. Models which match the plateau predict typical
depletion of factors of two. Fields et al. \cite{fields} argue that these
uncertainties represent additional errors of about a factor of two,
$\log(Li/H)\approx -9.8\pm 0.3$.
Even so, it does appear that a primordial abundance exists
in the range predicted by SBBN for the best estimates from the other
elements.

%(Recent results in three stars of the same mass, age and temperature
%seem to show unambiguous evidence of differences in popII lithium,
%apparently due to destruction by a factor of two or three.)

\section{Crisis or Concordance?}

The variety of data encourages a wide range of attitudes
in comparing with the Big Bang predictions.

The case for a ``crisis'' in Big Bang nucleosynthesis\cite{hat95}
is based on an
  upper bound  on $D/H$, leading to a
lower bound  on $\eta$ which conflicts with $Y_P$.
As explained above, this bound is not credible.
There is no Galactic conflict with a high deuterium abundance, since
chemical evolution can destroy D by a factor of 10 or more
and match metallicity vs age and other constraints, and the
evolution of $^3He$ is simply unknown.

But how good is the concordance, and how well constrained
is $\eta$?
A sensible and conservative view (\cite{cop95b},\cite{card96})  is that the 
 current spread of published
 results actually
reflects ignorance, or   ``systematic errors''. In this case,
one can say only that
all three elements are broadly consistent with $\eta $ over most of
 the range $10^{-9} $ to $10^{-10}$.
In spite of the concordance (granted, over a very wide
range in relative abundance),  this is a somewhat hollow victory
for SBBN.

Our goal however is more ambitious; we want a really precise
test of the model and a precise measurement of $\eta$.
If we put some faith in  our current best guesses,
  we can already take the value from 
the high $D/H$ quasar absorbers, and assign its errors
to the primordial abundance:  $(D/H)_p= 1.9\pm 0.5 \times 10^{-4}$.
This lets us estimate with some precision,
$\eta= 1.7\pm 0.3\times 10^{-10}$. Then we make a prediction
for helium, $Y_p= 0.233\pm 0.003 $ which is in astonishingly good
agreement with the best direct estimates
(assuming three light neutrino species as we expect), and a
similarly successful  prediction
for lithium, $\log(Li/H)=-9.75\pm 0.2$ (see e.g. \cite{dar}, \cite{fields}).
It could be that we are already
establishing concordance
 at a new level of precision. If so,  we   know the mean
baryon number density of the universe to $\pm 20\%$ accuracy.

Although it is premature to claim this 
precision just yet, 
many of the uncertainties in the current situation
will soon be settled by specific observations.  For example,
an ultraviolet Keck spectrum of Q1009+2956
would resolve many ambiguities in the determination of
low  $D/H$ in the $z=2.504$ absorber.
Better signal-to-noise at the blue end of the Q0014 spectrum
would help establish the Lyman series fit better and rule out
the interloper loophole there.
Correlation of $D/H$  with metals deserves close attention
as a potentially powerful constraint on chemical evolution
models, especially in the low redshift systems now
accessible with HST.
It would be fascinating to measure $D/H$ in a damped absorber;
this would be a different type of environment from those already measured,
closer to the conditions in the early Galaxy, and offers the possibility
of a very reliable measurement \cite{jen96}.

The helium measurements as well as the high deuterium
measurements seem to be pointing us in the direction of
low baryon density, $\eta_{10}\approx 2$,
$\Omega_b\approx 0.015$ for $h=0.7$. It was clear
at Princeton  that
such a low value
 is not particularly popular with model builders. One reason is
that it tends to imply a low density universe. For example, we can
use the cluster baryon fraction (using Steigman's equation 3)
 to infer that the density of dark matter is only $\Omega\approx 0.12$;
even allowing for systematic errors with cluster masses or baryon
segregation, this clearly implies an open universe or one dominated
by a cosmological constant.

We can also compare this density with an inventory of known
baryons in the universe. The accompanying table summarizes
estimates of the integrated density of known forms of baryons,
by integrating various mass functions based on systematic surveys
(e.g., \cite{fuk96},
\cite{fukagita},\cite{per92}). Because the dependence on $h$ varies, they are shown just
for $h=0.7$. The errors in the estimates are not well calibrated but
could easily be a factor of two for most components. The bottom line is
certainly consistent with a low baryon density, and indeed in this
case there is not a large reservoir of dark or unaccounted baryons.
If the baryon density is much higher, the extra baryons must be
hidden somewhere, which is increasingly difficult as constraints
improve on the density of compact objects (\cite{MACHO95}, \cite{MACHO96}) and
intergalactic gas.

 \bigskip

\centerline{Baryon Densities  for $h=0.7$}

\hbox to \hsize{\hss
\vbox{
\hrule height 0.6pt
\vskip 2pt
\hrule height 0.6pt
\halign{\strut
        #\quad\hfil & #\quad\hfil \cr
\noalign{\vskip 3pt}
 Form & $\quad\Omega _i$ \cr
\noalign{\vskip 3pt}
\noalign{\hrule height 0.6pt}
\noalign{\vskip 3pt}
Stars in spheroids & 0.0032 \cr
Stars in disks & 0.0017 \cr
Stars in irregulars & 0.0002 \cr
Neutral atomic gas (HI, HeI)  & 0.00033\cr
Molecular gas    & 0.0002\cr
Plasma in clusters  & 0.0026 \cr
Plasma in   groups & 0.0031 \cr
Cool intergalactic gas clouds & 0$.002$\cr
\noalign{\hrule height 0.6pt}
sum & 0.013  \cr
}
\vskip 3pt
\hrule height 0.6pt
\vskip 3pt
}
\hss}

\section{Acknowledgements}

I am grateful for many useful discussions with participants in
the 1996 workshop on Nucleosynthesis at the Institute for Nuclear
Theory in Seattle, funded by DOE.
 This work was supported at by NASA and the NSF
at the University of Washington.

\end{document}